\documentclass[preprint,aps,pra,showpacs,floatfix]{revtex4}
\usepackage{graphicx}
\usepackage{times}
\usepackage {euscript}
\usepackage{amsmath}
\usepackage{amsthm}
\usepackage{epsf}
\usepackage{epsfig}
\renewcommand{\vec}[1]{{\mbox{\boldmath$#1$}}}

\makeatletter  
\def\p@subsection{}  
\makeatother  
\begin{document}  
\title{  
Relativistic calculations of the x-ray emission   
following the Xe-Bi$^{83+}$ collision  
}  
%
%
\author{ Y.~S.~Kozhedub,$^{1,2}$ V.~M.~Shabaev,$^1$ I.~I.~Tupitsyn,$^1$   
A.~Gumberidze,$^{3}$ S.~Hagmann,$^{3}$ G.~Plunien$^4$, and 
Th.~St\"ohlker$^{3,5}$  
}  
\affiliation{  
$^1$  
Department of Physics, St. Petersburg State University,   
Ulianovskaya~$1$,  
Petrodvorets, St.~Petersburg 198504, Russia \\  
$^2$  
SSC RF ITEP of NRC "Kurchatov Institute",  
Bolshaya Cheremushkinskaya~25, Moscow, 117218, Russia\\  
$^3$  
GSI Helmholtzzentrum f\"ur Schwerionenforschung, 
64291 Darmstadt, Germany \\  
$^4$  
Institut f\"ur Theoretische Physik, Technische Universit\"at Dresden,  
Mommsenstra{\ss}e 13, D-01062 Dresden, Germany \\  
$^5$  
Helmholtz-Institut Jena, D-07743 Jena, Germany \\  
}  
%
\begin{abstract}  
We study the x-ray emission following the collision of a  Bi$^{83+}$ ion  
with a neutral Xe atom at the projectile energy 70 MeV/u.  
The collisional and post-collisional processes are treated separately.   
The probabilities of various many-electron processes at the collision are 
calculated   
within a relativistic independent electron model using the  
coupled-channel approach with atomic-like Dirac-Fock-Sturm orbitals.   
The analysis of the post-collisional processes resulting in the x-ray 
emission is based on the fluorescence yields, the radiation and Auger decay 
rates,  and allows to derive intensities of the x-ray emission and compare them 
with experimental data. A reasonable agreement between   
the theoretical results and the recent experimental data   
is observed.  The role of the relativistic effects is investigated.  
\end{abstract}

\maketitle  
%
%
\section{Introduction}  
 
Collisions between multiply charged ions and atoms have been  
a subject of intensive studies for many years.  
Depending on the number of active electrons in the target atom and the charge   
state of the projectile, the multiple electronic ionization, excitation and 
capture   
are possible and multiple excited states can be formed.   
These states decay by emission of photons and/or  
electrons (Auger decay), both carrying information on the collision dynamics.   
 
A significant progress in  
describing the quantum dynamics  
of electrons in low-energy ion-atom collisions  
has been made by studying various collision systems experimentally and   
theoretically 
(see Refs.~\cite{Schuch:prl:1979,Liesen:prl:1980,
Hagmann:pra:1982,Tanis:prl:1982,Pepmiller:pra:1985,
Fritsch:pr:1991,Barat:jpb:1992,Bransden_92,Kurpick:pra:1995,
Ludde:jpb:1996,Arnau:scr:1997,Otranto:pra:2006, 
Dennerl:scr:2010,Tolstikhina:pu:2013} 
and references therein). 
The results are especially successful for low-$Z$   
($Z$ is the nuclear charge number) atoms/ions  with a small number of active 
electrons.  
However, for collisions between highly charged ions (HCI) and heavy  
targets, both experimental data and theoretical analyses are  
scarce due to the complex nature of the problem.  
Meanwhile collisions of HCI provide a unique tool  
for tests of relativistic and quantum electrodynamics (QED) effects  
in the scattering processes~\cite{Eichler_95,Shabaev_02,Eichler_07}.  
Investigations of such processes can also give an access  
to  QED in supercritical  fields, provided the total charge of the colliding  
nuclei is larger than the critical one, $Z_c=173$ (see, e.g.,   
Ref.~\cite{Greiner_85} and references therein).  
 
Experimental investigations aimed at the comprehensive   
study of various processes in low-energy heavy ion-atom collisions  
are anticipated at GSI and FAIR facilities   
(Darmstadt, 
Germany)~\cite{FAIR,SPARC,CRYRING,Hagmann_2012,Gumberidze_2012_prop}. In a 
recent experiment \cite{Gumberidze:2011},    
the intensities of the post-collisional x-ray emissions have been   
resolved for collisions of   
Bi$^{83+}$ ions with Xe target atoms at the projectile energy~$70$~MeV/u.  
These experiments require the corresponding theoretical calculations.  
In the present paper, we perform a relativistic quantum-mechanical  
calculation of the Bi$^{83+}$-Xe  collision within an independent  
electron model using   
the coupled-channel approach with atomic-like Dirac-Fock-Sturm   
orbitals~\cite{Tupitsyn_12,Kozhedub:ps:2013}. The post-collision decay   
analysis allows us to derive the x-ray radiation intensities and compare them   
with the experimental data. We also study  the role of the   
relativistic effects.  
 
The paper is organized as follows.  
In section~\ref{subsec:Collision1},   
the theoretical approach used to calculate the collision  
dynamics is briefly described.  
Section~\ref{subsec:radiation} is devoted to the description of  
post-collision processes coming to the X-ray radiation.   
In section~\ref{subsec:Collision2}, the results for the  
single and multiple electronic dynamic  
probabilities are given, while in  
section~\ref{subsec:radiation2} the theoretical  
x-ray intensities are presented and compared  
with the experimental data.  
Some general remarks and comments are given in the last section.  
 
Throughout the paper atomic units ($\hbar=e=m_e=1$) are   
used.  
%
\section{Theory}  
\label{sec:theory}  
We consider the collision of a bare bismuth (Bi$^{83+}$) with a   
target xenon atom at the projectile energy~$E=70$~MeV/u.   
At this energy the ionization of the target is the dominant   
process, but the target excitation and the electron capture by the projectile   
ion are also possible.   
The excited xenon and bismuth ions decay via Auger processes or radiatively.  
In any case, the de-excitation processes (Auger  or radiative decays)  
take much more time   
than the electronic dynamic processes during the collision.   
Therefore, the collisional and  the post-collisional decay dynamics can be   
viewed as being  independent from each other and thus we can treat them  
separately. For the collision part, we solve the time-dependent Dirac   
equation within the semi-classical approximation and an independent electron   
model. The corresponding theory is presented in   
section~\ref{subsec:Collision1}. In section~\ref{subsec:radiation}, the x-ray   
radiation intensities for the post-collision part are evaluated using an   
analysis based on the fluorescence yields, the radiation and Auger decay   
rates.

\subsection{Collision}  
\label{subsec:Collision1}  
%
 
The ion-atom collision is considered in the semiclassical approximation, where  
the atomic nuclei move along the classical trajectories. The nuclei  are   
considered as sources of a time-dependent external potential, whereas the   
electrons are treated quantum-mechanically.  
The many-electron time-dependent Dirac equation  
is considered in the framework of an independent particle model, in which the   
many-particle Hamiltonian $\hat H$ is approximated by a sum of single-particle  
Hamiltonians, $\hat H^{\rm eff}=\sum_{i} \hat h^{\rm eff}_{i}$, reducing the   
many-electron problem to a set of single-particle Dirac equations:  
\begin{equation}  
i \frac{\partial \psi_i(t)}{\partial t} = \hat h^{\rm eff}_{i}(t) \,  
\psi_i(t) \quad {\rm with} \quad i=1,\ldots,N,  
\label{eq:Dirac}  
\end{equation}  
where the wave functions $\psi_i(t)$ have to satisfy the initial conditions for 
the $N$ electrons:  
\begin{equation}  
\lim_{t\rightarrow -\infty}(\psi_i(t) - \psi_{i}^{0}(t))=0  
\quad {\rm with} \quad i=1,\ldots,N.  
\label{eq:}  
\end{equation}  
The many-electron wave function is given by a Slater determinant made-up from   
the single-particle wave functions.   
The two-center Dirac-Kohn-Sham Hamiltonian is used for $\hat h^{\rm eff}$:  
\begin{equation}  
\hat h^{\rm eff} =c (\vec{\alpha} \cdot \vec{p}) +  
\beta \, c^2 +   
V_{\rm nucl}^{A}(\vec{r}_A) + V_{\rm nucl}^{B}(\vec{r}_B) +  
V_C[\rho] + V_{xc}[\rho]  
\,,  
\end{equation}  
where $c$ is the speed of light and $\vec{\alpha}$, $\beta$ are the  
Dirac matrices.   
$V_{\rm nucl}^{\alpha}(\vec{r}_\alpha)$ and  
$V_C[\rho] = \int \, d^3\vec{r^{\prime}} \,   
\frac{\rho(\vec{r}^{\prime})}{|\vec{r}-\vec{r}^{\prime}|}$  
are the electron-nucleus interaction and the electron-electron Coulomb  
repulsion potentials, respectively,  
and   
$\rho(\vec{r})$ is the electron density of the system.   
The exchange-correlation  
potential $V_{xc}[\rho]$ was taken in the Perdew-Zunger  
parametrization~\cite{Perdew_81} including the self-interaction correction.

 
To solve Eq.~(\ref{eq:Dirac})   
we use the coupled-channel approach with atomic-like Dirac-Fock-Sturm orbitals  
$\varphi_{\alpha,a}$~\cite{Tupitsyn_12,Bratsev_77}, localized at   
the ions  
(atoms). The time-dependent single-particle wave function   
$\psi_i(t)$  is represented as   
\begin{equation}  
\psi_i(\vec{r},t)  =  \displaystyle \sum_{\alpha=A,B} \, \sum_{a}  
C^{i}_{\alpha, a}(t) \, \varphi_{\alpha,a} (\vec{r}_\alpha(t)) \,.  
\label{eq:expan1}  
\end{equation}  
Here index $\alpha=A,B$ labels the centers, index $a$  
enumerates basis functions at the given center,   
$\vec{r}_{\alpha}=\vec{r}-\vec{R}_\alpha$,  
and  
$\varphi_{\alpha,a} (r_\alpha)$ is the central-field  
bispinor centered at the point $\alpha$.  
The insertion of the  
expansion~(\ref{eq:expan1})  
into the Dirac-Kohn-Sham equations~(\ref{eq:Dirac})    
leads to the  
well-known coupled-channel equations for the coefficients   
$C^{i}_{\alpha, a}(t)$  
\begin{equation}  
i \sum_{\beta, b} \langle \varphi_{\beta, b} \mid \varphi_{\alpha, a} \rangle   
\frac{dC^{i}_{\beta, b}(t)}{d t}  =   
\sum_{\beta, b} \,  
\langle \varphi_{\alpha, a} \mid (\hat h^{\rm eff}_i   
- i \frac{\partial}{\partial t})\mid \varphi_{\beta,b}  
\rangle \,   
C^{i}_{\beta, b}(t)  \,,  
\label{eq:cap_chan}  
\end{equation}  
where indices $\alpha, a$ and $\beta, b$ enumerate the basis functions of both  
centers.  
 
The expansion coefficients are determined employing the direct evolution  
(exponential) operator method~\cite{Tupitsyn_10}, which is more stable  
compared to the others, such as, e.g., the Crank-Nicholsen propagation  
scheme~\cite{Crank_47} and the split-operator method~\cite{Fett_82}.  
To obtain the matrix representation of the exponential operator  
in the finite basis set one has to diagonalize the generalized complex  
Hamiltonian matrix at each time step. Since our basis set is not too large,  
the diagonalization procedure is not too time consuming.

The transition probability from a given  
initial configuration of colliding ions $(i_1,\ldots,i_N)$ to a given final   
configuration $(f_1,\ldots,f_N)$ is defined by   
\begin{align}  
P_{f_1,\ldots,f_N}=  
| \langle  f_1,\ldots,f_N  
|  i_1,\ldots,i_N; t_f\rangle  
|^2  
.  
\end{align}  
Here, $|i_1,\ldots,i_N;t \rangle$ denotes   
the Slater determinant constructed from the solutions of the effective   
one-particle   
equations~(\ref{eq:Dirac})  at time $t$ for the initial condition   
$(i_1,\ldots,i_N)$.  
The probability can be written as the $(N \times N)$ determinant of the   
one-particle   
density matrix~\cite{Ludde:jpb:1985,Kurpick:cpc:1993}  
\begin{align}  
P_{f_1,\ldots,f_N}= \rm{det}(\gamma_{nn'}),  
\quad n,n'=1,\ldots,N,  
\end{align}  
where  
%
\begin{align}  
\gamma_{nn'} =   
\langle f_n | [  
\sum^{N}_{i=1} |\psi_i(x,t=\infty)\rangle \langle \psi_i(x',t=\infty) |  
] | f_{n'} \rangle .  
\end{align}  
The probability to find $q$ electrons in given states is determined by   
formulas  
\begin{equation}  
P_{f_1,\ldots,f_q}= \sum_{f_{q+1}<\cdots<f_N} P_{f_1,\ldots,f_N}\,,   
\quad q<N ,  
\end{equation}  
\begin{equation}  
P_{f_1,\ldots,f_q}=    
\rm{det}(\gamma_{nn'}),  
\quad n,n'=1,\ldots,q,  
\quad q<N .  
\end{equation}  
The corresponding inclusive probability for a configuration with $q$   
occupancies  and $L-q$ holes, in terms of inclusive probabilities related only  
to occupancies, takes the form   
\begin{equation}  
P_{f_1,\ldots,f_q}^{f_{q+1},\ldots,f_L}=    
P_{f_1,\ldots,f_q}   
-\sum_{f_{q+1}}  P_{f_1,\ldots,f_q,f_{q+1}}  
+\sum_{f_{q+1}<f_{q+2}}  P_{f_1,\ldots,f_q,f_{q+1},f_{q+2}}  
+\cdots+  
(-1)^{L-q}P_{f_1,\ldots,f_q,f_{q+1},\ldots,f_{L}}\,.  
\end{equation}  
%
 
\subsection{X-ray radiation}  
\label{subsec:radiation}

In order to evaluate the intensities of the K and L x-ray radiation of the ions 
after the collision we use a scheme, the main steps of which are the following: 
 
(1) to look for states $f$ of the ions which can de-excite via the   
considered x-ray radiation;  
 
(2) to calculate the probabilities $P_f$ to find the  
system in the states $f$ after the collision;  
 
(3) to determine the radiative de-excitation probabilities  
with $m$ emitted x-ray photons  
for the states $f$ under consideration, $P^{\rm{rad}}_{m}(f)$;  
 
(4) to evaluate the "relative" x-ray radiation intensities  
(the number of the emitted photons per collision) as  
$I=\sum_{f} m P^{\rm{rad}}_{m}(f) P_f $.  
 
The values $P_f$ are derived as described in section~\ref{subsec:Collision1}.   
To determine the values of  $P^{\rm{rad}}_{m}(f)$ we are guided   
by the fluorescence yield, the radiation and Auger decay rates, and   
some features of the ion states $f$.

As a brief example, let us consider the K x-ray radiation of the xenon ion   
following the collision. The K radiation is possible for states having the 
K-shell vacancies. Actually there are two different cases: the states   
with  only one and the states with exactly two K-shell vacancies. These states  
are filled through the radiative de-excitation, which probability is defined by 
the fluorescence yield coefficient for the xenon K shell $P^{\rm{rad}}(\rm{K})$ 
(it is assumed that the higher energy shells are occupied rather densely).   
If $P_1$ and $P_2$ are the probabilities to find one and two K-shell   
vacancies, the relative K x-ray radiation intensity is  
$I_{\rm{K}}=P^{\rm{rad}}(\rm{K})(P_1 + 2 P_2)$.

\section{Results of the calculations and discussion}  
\label{subsec:results}  
 
\subsection{Collision}  
\label{subsec:Collision2}  
%
The method described in section~\ref{subsec:Collision1}  
was applied to the Xe and Bi$^{83}$ collision at the projectile energy   
$70$~keV/u.  
Since we are interested in the K and L x-ray radiation of the ions,  
we focus first on the study of the K-, L-shell electronic population    
probabilities for the colliding ions.  
 
\begin{figure}[hbt]  
\centering  
\vspace{-0.2cm}  
\includegraphics[width=10.5cm,clip]{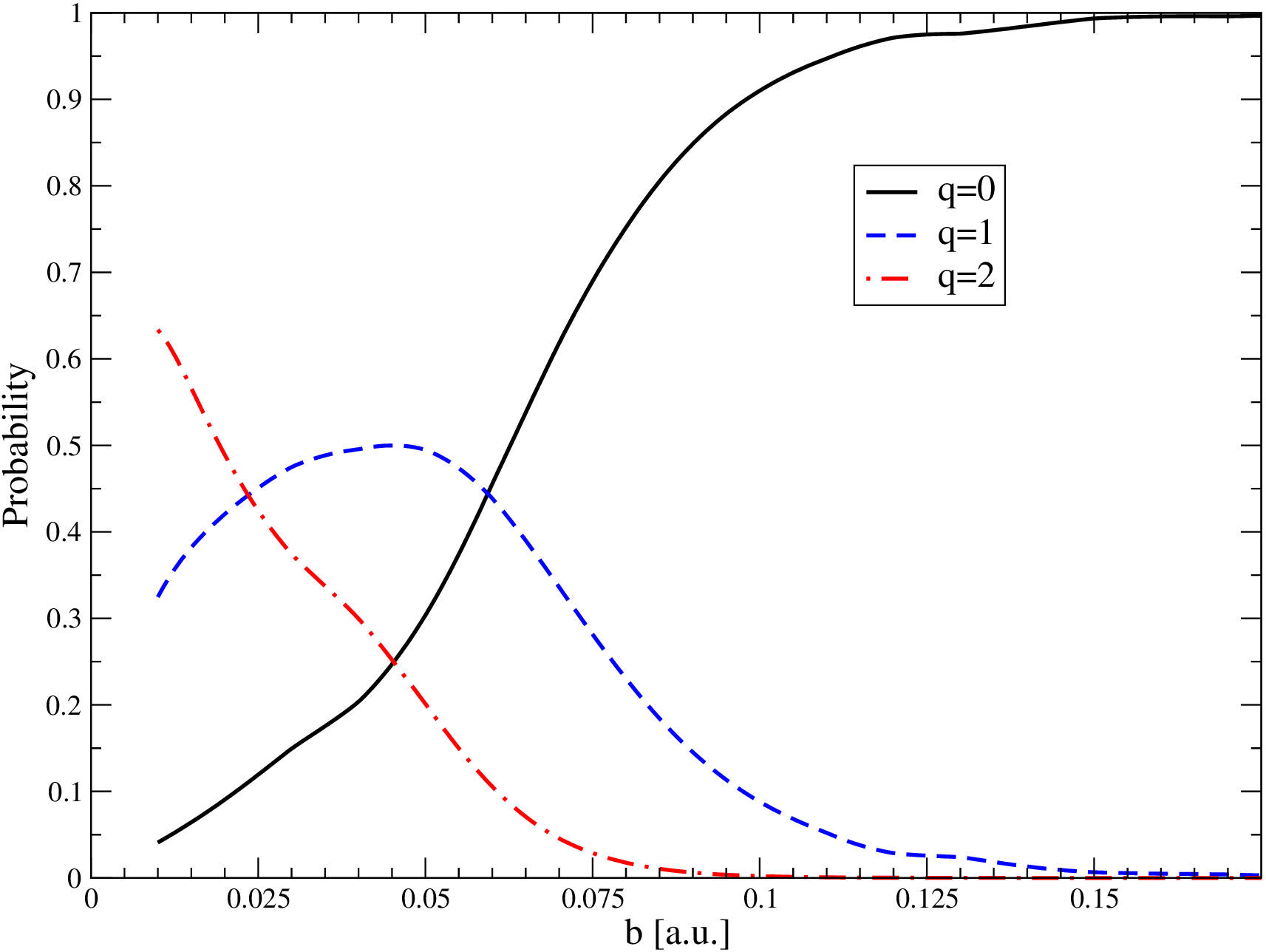}  
\caption{  
The probabilities of the Xe $q$-K-shell-vacancy production   
in the Xe-Bi$^{83+}$ collision as functions of the impact parameter $b$.}  
\label{Fig:Xe_Kvac}  
\end{figure}

\begin{figure}[hbt]  
\centering  
\vspace{-0.2cm}  
\includegraphics[width=10.5cm,clip]{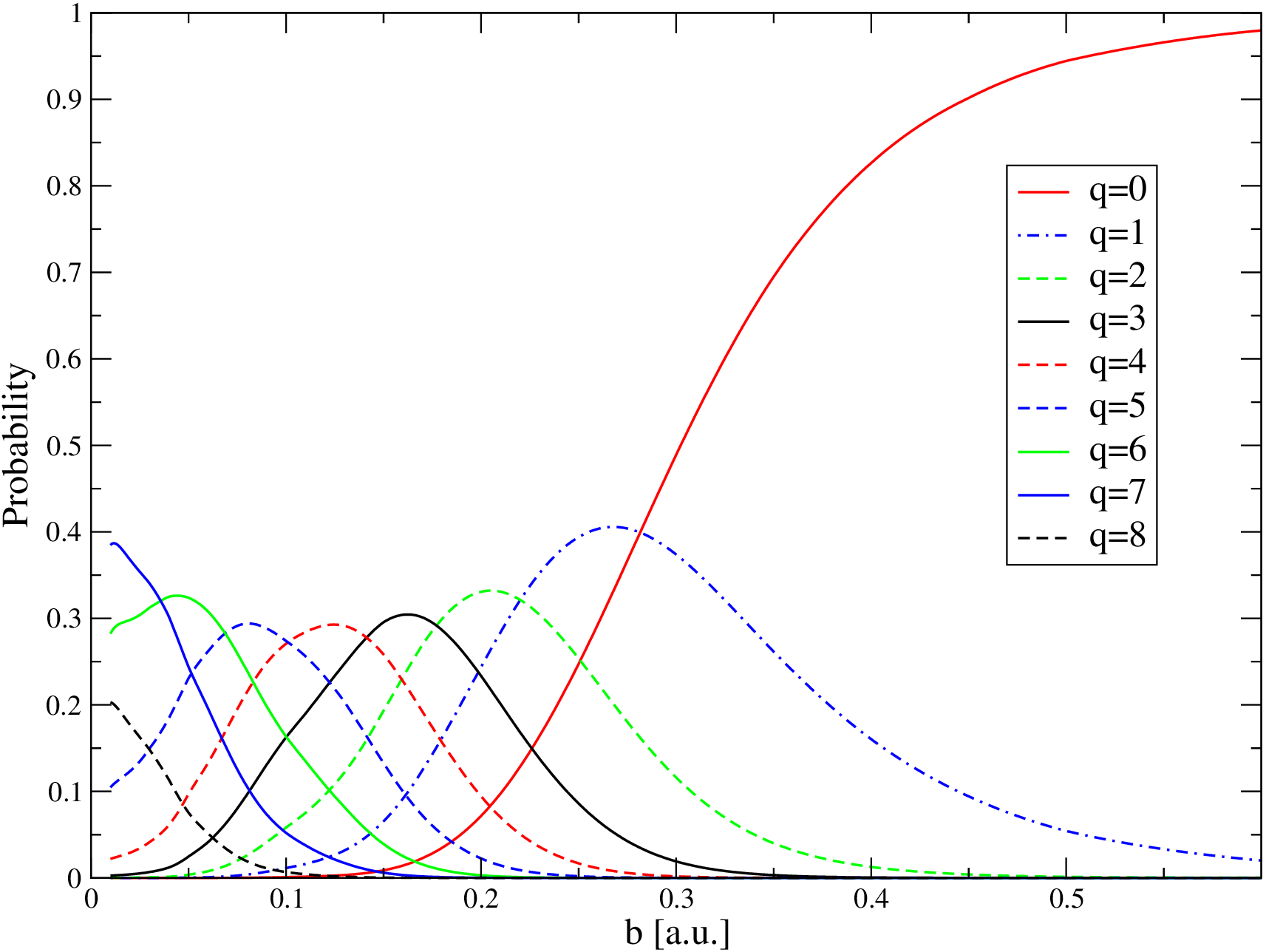}  
\caption{  
The probabilities of the Xe $q$-L-shell-vacancy production   
in the Xe-Bi$^{83+}$ collision as functions of the impact parameter $b$.  
}  
\label{Fig:Xe_Lvac}  
\end{figure}

Figures~\ref{Fig:Xe_Kvac} and~\ref{Fig:Xe_Lvac} show the probabilities   
of the $q$-vacancies creation in the K and L shells of the target ion (xenon),  
correspondingly, as functions of the impact parameter.   
We note that the probability of the K-shell-vacancy production becomes   
significant for the impact parameter~$b$ less than $0.1$~a.u.   
and for $b<0.06$~a.u. this probability is dominating (the size of the K shell   
is about $0.02$~a.u.). When the impact parameter is close to zero, the   
probability to find at least one electron in the K shell is almost vanishing.

As one can see from Fig.~\ref{Fig:Xe_Lvac},   
the results for the L-shell-vacancy production probabilities are   
similar to the K shell ones. The L-shell electron loss probability grows   
rapidly when the impact parameter $b$ becomes smaller than $0.5$~a.u. (the size  
of the L shell is about $0.1$~a.u.). At $b<0.3$~a.u. the vacancy creation is 
the dominant process and, moreover, at  $b<0.2$~a.u.  
the multiple vacancy production takes mainly place.

%
 
%
 
\begin{figure}[hbt]  
\centering  
\vspace{-0.2cm}  
\includegraphics[width=10.5cm,clip]{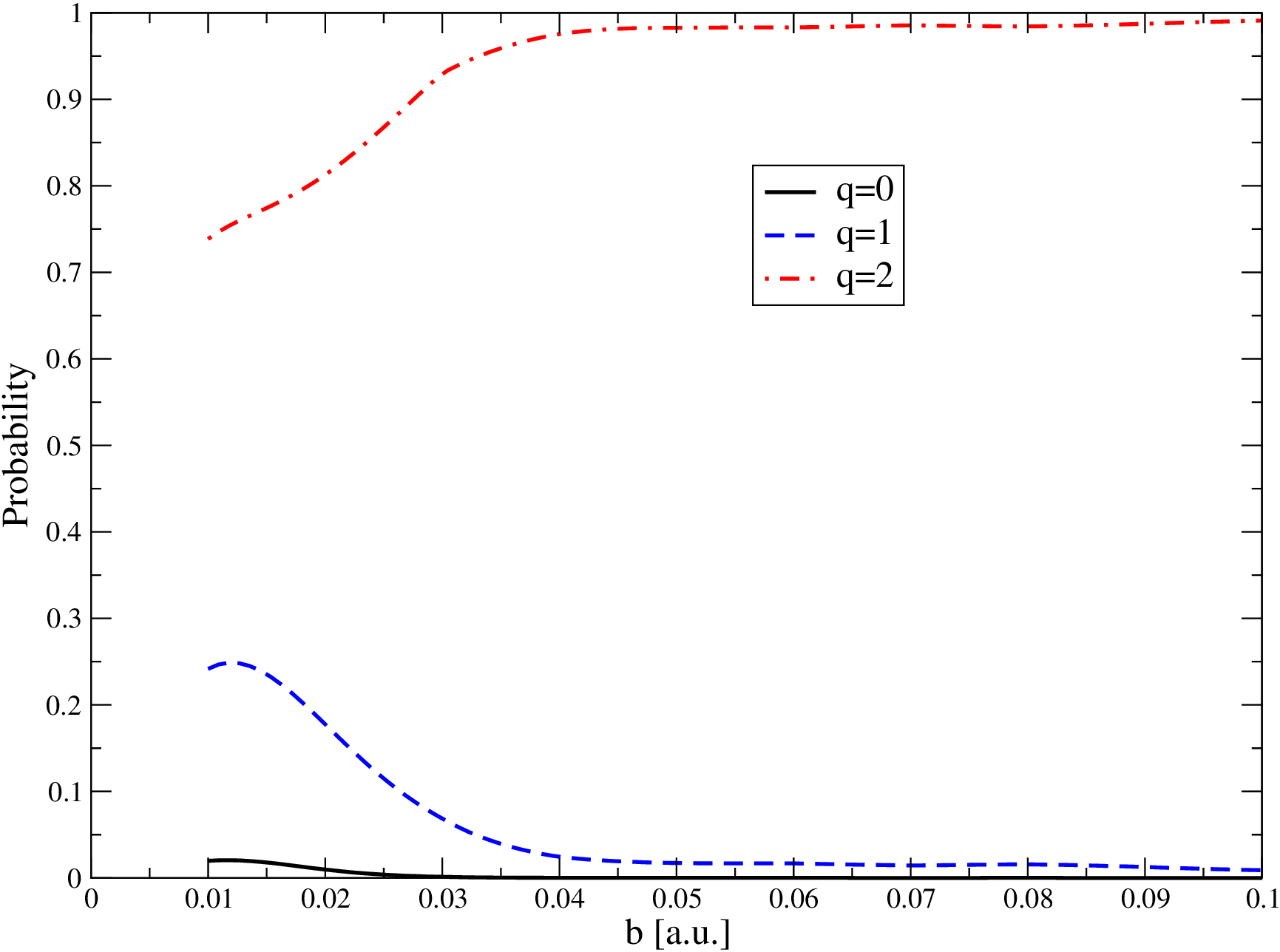}  
\caption{The probabilities of the Bi $q$-K-shell-vacancy surviving   
in the Xe-Bi$^{83+}$ collision as functions of the impact parameter $b$.  
}  
\label{Fig:Bi_Kvac}  
\end{figure}

\begin{figure}[hbt]  
\centering  
\vspace{-0.2cm}  
\includegraphics[width=10.5cm,clip]{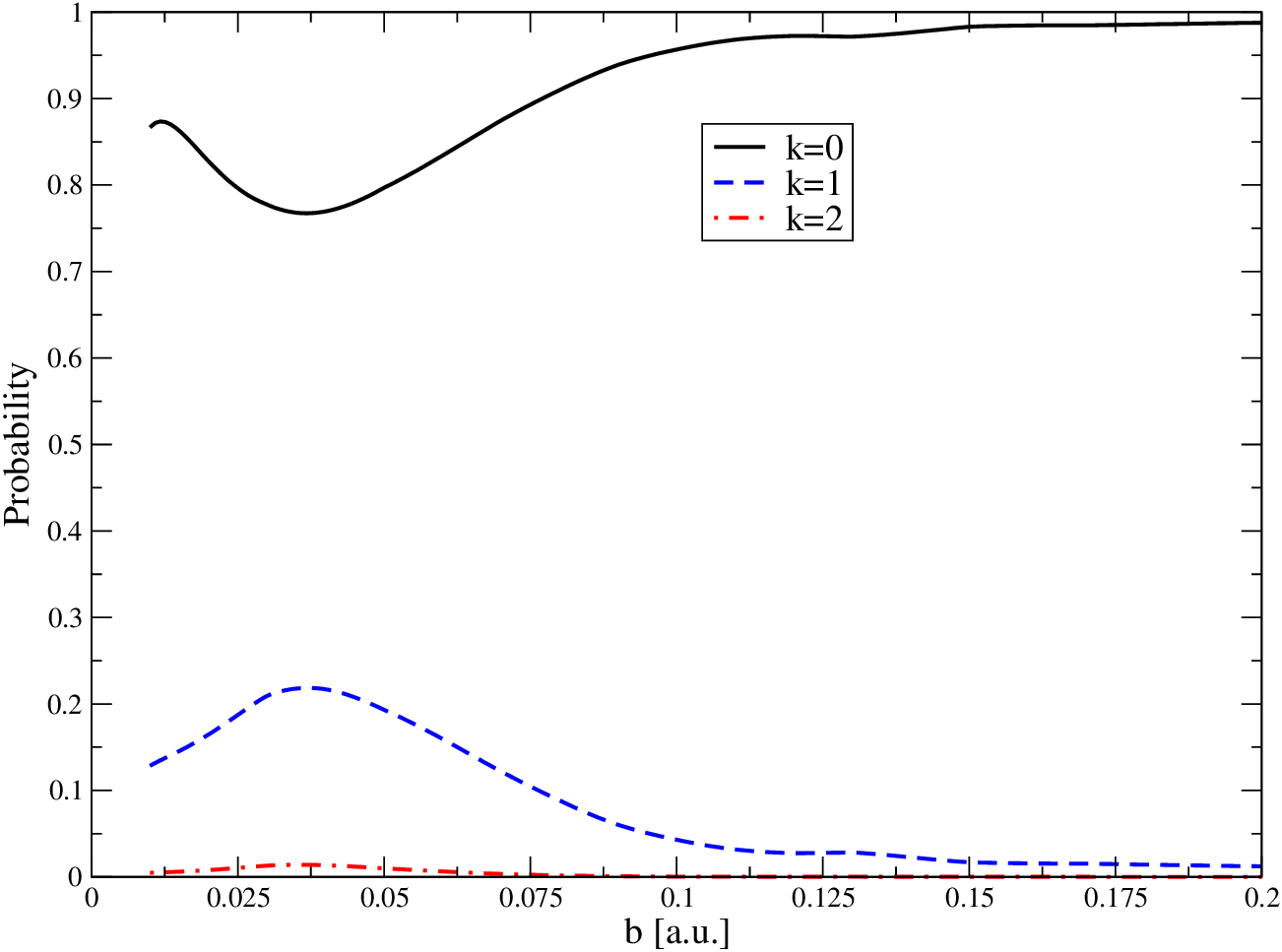}  
\caption{  
The probabilities of the Bi $k$-L$_1$-shell~$(2s)$ electronic population  
in the Xe-Bi$^{83+}$ collision as functions of the impact parameter $b$.%
}  
\label{Fig:Bi_L1}  
\end{figure}

\begin{figure}[hbt]  
\centering  
\vspace{-0.2cm}  
\includegraphics[width=10.5cm,clip]{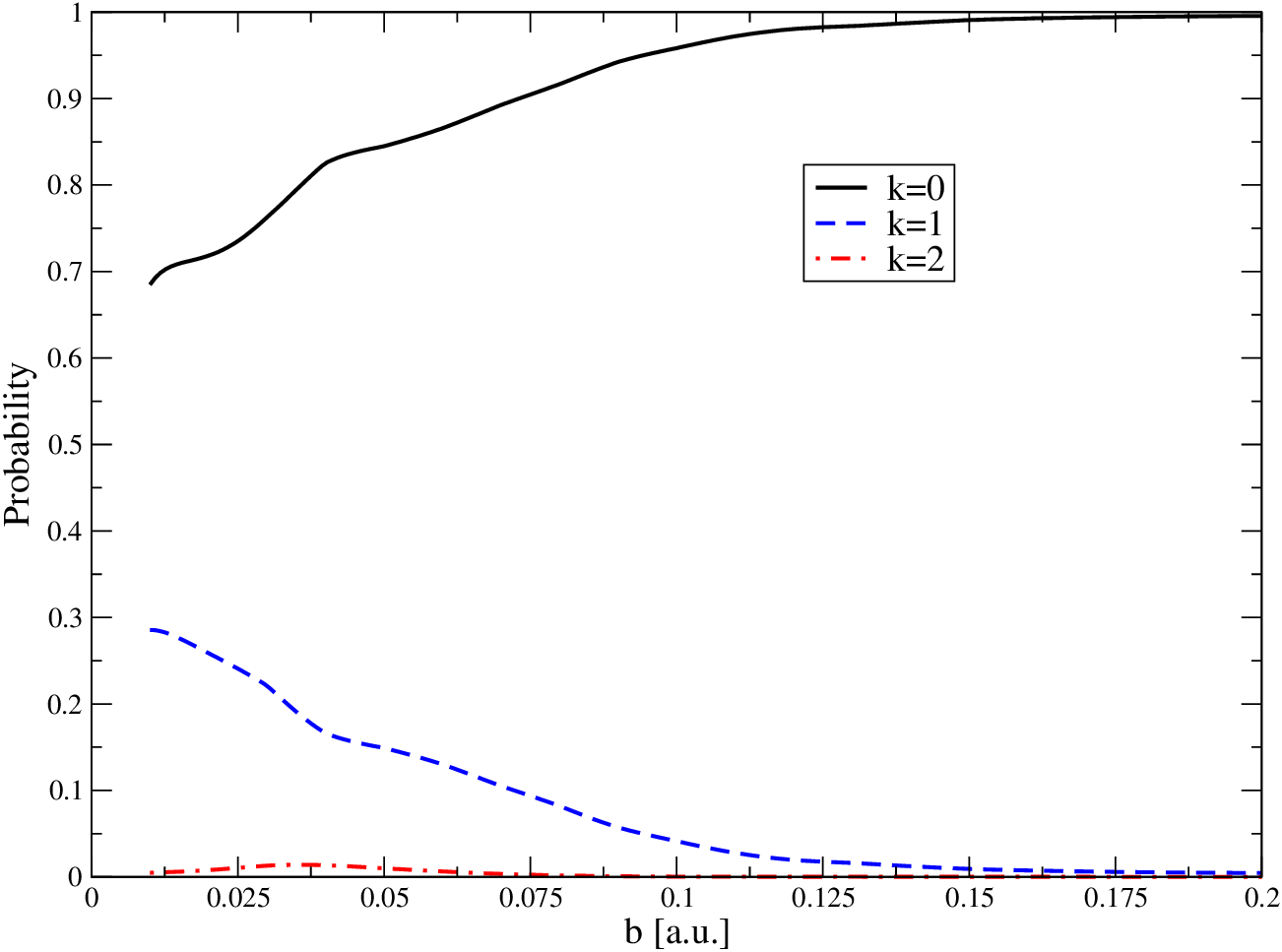}  
\caption{  
The probabilities of the Bi $k$-L$_2$-shell~$(2p_{1/2})$ electronic population  
in the Xe-Bi$^{83+}$ collision as functions of the impact parameter $b$.%
}  
\label{Fig:Bi_L2}  
\end{figure}

\begin{figure}[hbt]  
\centering  
\vspace{-0.2cm}  
\includegraphics[width=10.5cm,clip]{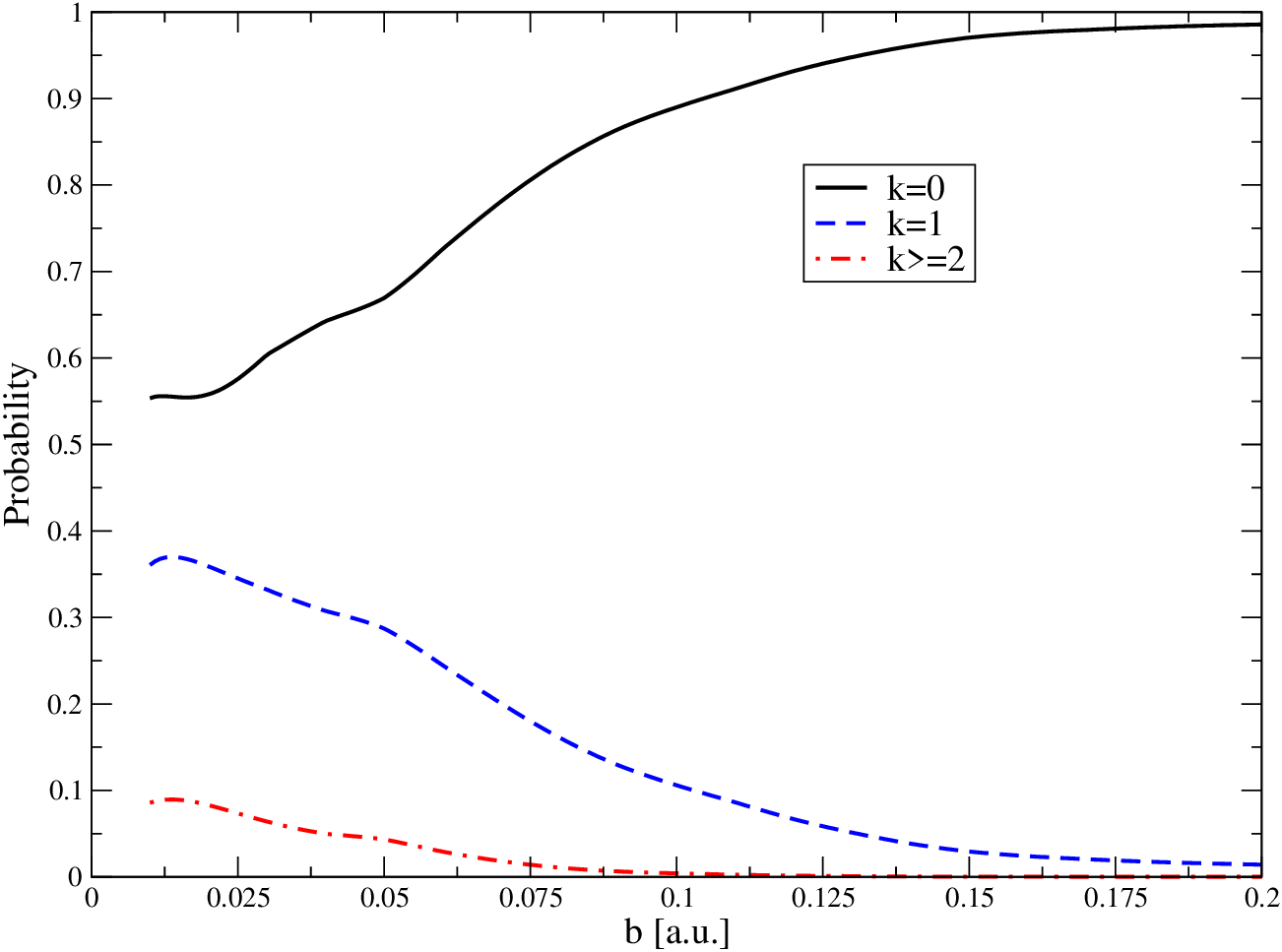}  
\caption{  
The probabilities of the Bi $k$-L$_3$-shell~$(2p_{3/2})$ electronic population  
in the Xe-Bi$^{83+}$ collision as functions of the impact parameter $b$.%
}  
\label{Fig:Bi_L3}  
\end{figure}

The results of the calculations for the post-collisional bismuth states   
are presented in Figures~\ref{Fig:Bi_Kvac}-\ref{Fig:Bi_L3}.  
%
Fig.~\ref{Fig:Bi_Kvac} displays that the K shell of bismuth is almost empty   
for a wide range of the impact parameter value. The maximum of the probability 
to  observe  at least one electron in the K shell achieves $0.25$ at 
$b=0.012$~a.u. (the size of the K shell is about $0.016$~a.u.).  
 
%
 
%
 
%
 
The L bismuth shells are conveniently described in terms of   
electronic population probabilities. The $k$-L-shell electronic population   
probabilities for the L$_{1}$~$(2s)$, L$_{2}$~$(2p_{1/2})$,   
L$_{3}$~$(2p_{3/2})$ shells as functions of the impact parameter are presented  
in Figures~\ref{Fig:Bi_L1}, \ref{Fig:Bi_L2}, and~\ref{Fig:Bi_L3},   
respectively.  
The probabilities are not large and we can conclude that the probability to   
find more than two electrons at the L shell is almost zero.  
 
\subsection{X-ray radiation}  
\label{subsec:radiation2}  
 
\begin{figure}  
\centering  
\vspace{-0.2cm}  
\includegraphics[width=10.5cm,clip]{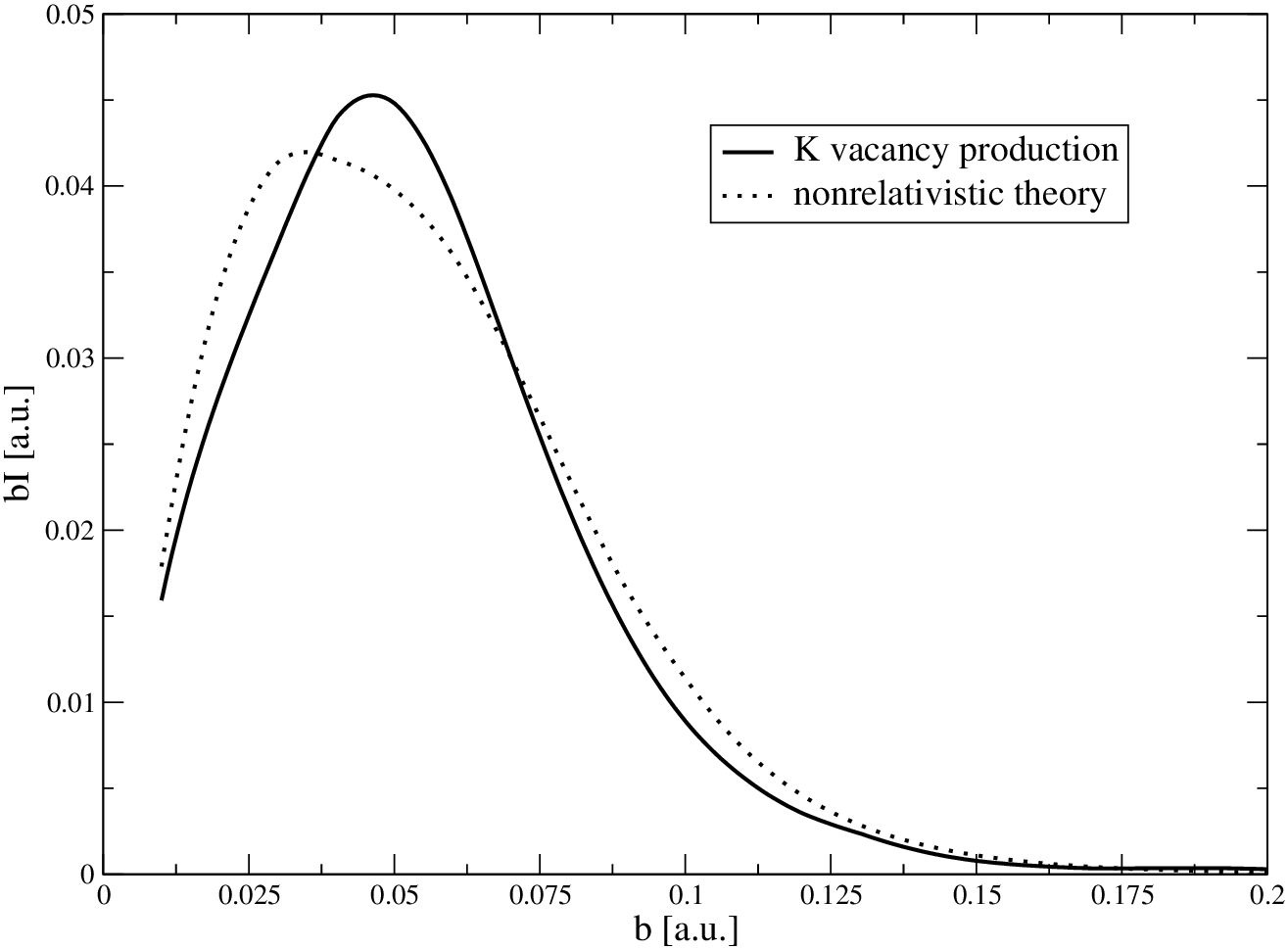}  
\caption{The intensity $I$ of the Xe K-shell-vacancy production   
weighted by the impact parameter in the Xe-Bi$^{83+}$ collision as a function   
of the impact parameter $b$. The dotted line indicates the results of the  
nonrelativistic calculations.}  
\label{Fig:Xe_Krad}  
\end{figure}

\begin{figure}  
\centering  
\vspace{-0.2cm}  
\includegraphics[width=10.5cm,clip]{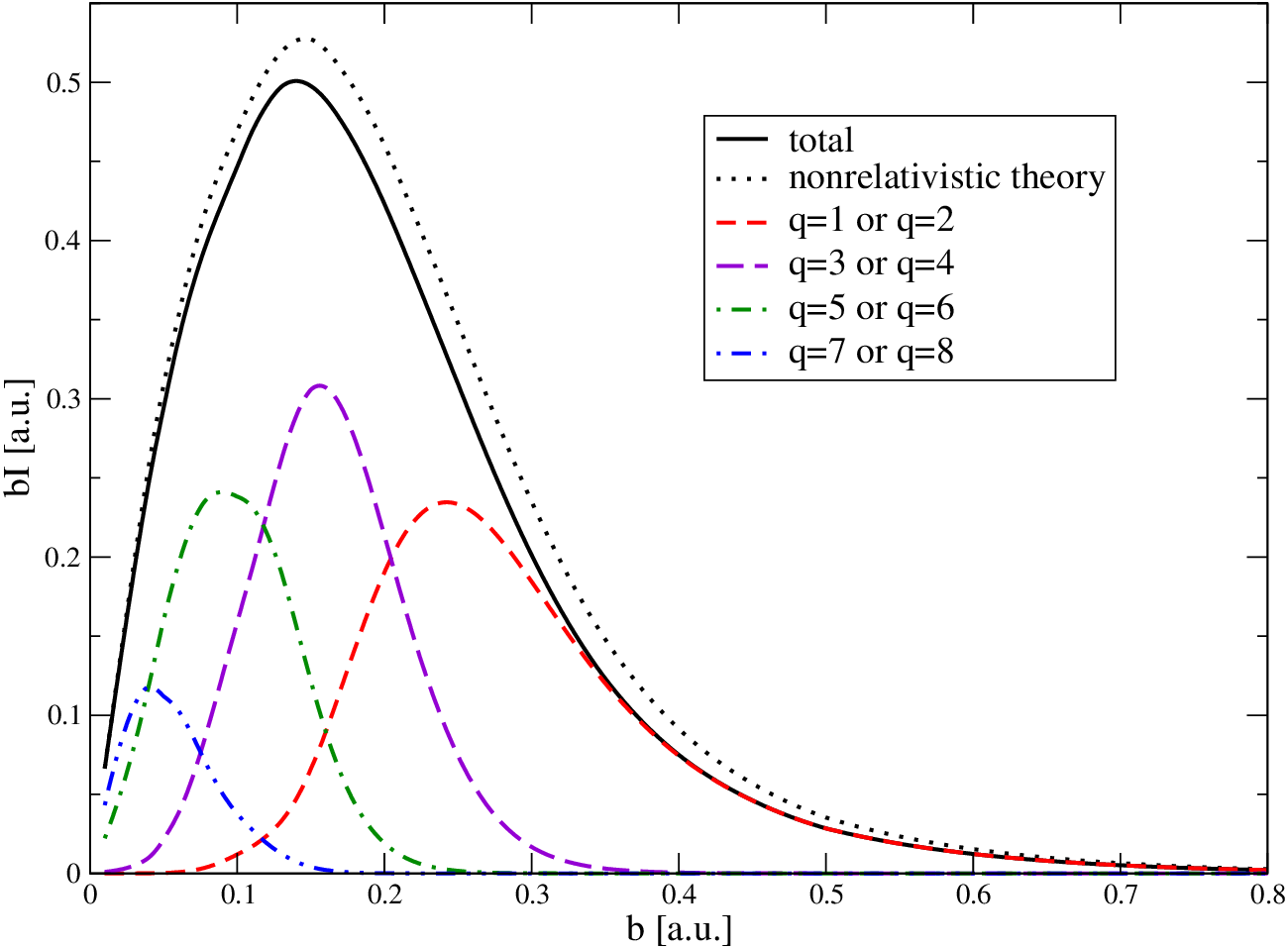}  
\caption{The total- and $q$-intensities $I$ of the Xe L-shell-vacancy   
production weighted by the impact parameter in the Xe-Bi$^{83+}$ collision as   
functions of the impact parameter $b$. The dotted line indicates the   
total results of the nonrelativistic calculations.}  
\label{Fig:Xe_Lrad}  
\end{figure}

After the collision the xenon atom (target) is generally a highly ionized and   
excited ion with a large number of the K- and L-shell vacancies. Nevertheless,  
the binding electrons are in plenty and during de-excitation processes all   
the K- and L- shell vacancies are filled by electrons   
with the emission of the K and L x-ray photons in accordance with  
the fluorescence yield probabilities   
$P^{\rm{rad}}(\rm{K})$ and $P^{\rm{rad}}(\rm{L})$,   
correspondingly.  
Let $I_{\rm{K}}(b)=\sum_{q} q P_{\rm{K},q}(b)$ is the intensity of the   
K-shell-vacancy production per collision as a function of the impact parameter  
$b$.  
Here $P_{\rm{K},q}(b)$ is the probability of the $q$-K-shell-vacancy   
production. Then the intensities of the K and L x-ray radiation are given by  
\begin{equation}  
I_{\rm{K}}^{\rm{rad}}=2\pi P^{\rm{rad}}({\rm{K}})
\int_{0}^{\infty} I_{\rm{K}}(b) b db, 
\label{eq:I_rad_K}  
\end{equation}  
\begin{equation}  
I_{\rm{L}}^{\rm{rad}}=2\pi P^{\rm{rad}}({\rm{L}})   
\int_{0}^{\infty} I_{\rm{L}}(b) b db.  
\end{equation}  
The results of the calculations for $I_{\rm{K}}(b)$ and $I_{\rm{L}}(b)$   
weighted by the impact parameter are presented in Figures~\ref{Fig:Xe_Krad}   
and~\ref{Fig:Xe_Lrad}, respectively.   
To investigate the role of the relativistic effects we performed the same   
calculations in the nonrelativistic limit by multiplying the standard value of  
the speed of light by the factor $1000$.   
In the figures the obtained nonrelativistic data are indicated by the dotted 
lines. In Fig.~\ref{Fig:Xe_Lrad}, the contributions from   
the $q$-L-shell-vacancy production processes to the total intensity are also   
presented.  
For the evaluation of $I_{\rm{K}}^{\rm{rad}}$ and $I_{\rm{L}}^{\rm{rad}}$ the   
values of the fluorescence yield $P^{\rm{rad}}(\rm{K})=0.89(1)$ and   
$P^{\rm{rad}}(\rm{L})=0.09(1)$   
are taken from Refs~\cite{Grieken:2002,Hubbel:1994}. These values are based on  
the experimental and theoretical analysis.   
 
We note that at the values of the impact parameter, where the   
integral~(\ref{eq:I_rad_K}) is assembled (see Fig.~\ref{Fig:Xe_Krad}),  
the L shell is almost empty (see Fig.~\ref{Fig:Xe_Lvac}). It decreases the   
probability of the K shell filling by the Auger decay and, therefore, the value  
of $P^{\rm{rad}}(\rm{K})$ is underestimated. Thus, it is reasonable to   
expect a little bit underestimated (up to $5\%$) the value of   
$I_{\rm{K}}^{\rm{rad}}$. The fluorescence yield coefficient for the   
L shell is actually individual for each L subshell (L$_1$, L$_2$, L$_3$). Here 
we use the average value of the coefficient over the subshells. With a high   
probability (about~$70\%$) a vacancy in the K shell leads to a   
vacancy in the L shell. This is in  accordance with the experimental data  
for the relative intensities of the K$_\alpha$ and K$_\beta$   
x-ray lines~\cite{Gumberidze:2011}. But the  
intensity of the K-shell-vacancy production does not exceed $2\%$ of the L   
shell one. That is why we neglect this contribution to the L-shell-vacancy   
production processes.  
The final values of the radiative intensities per collision or the total cross  
sections of the radiation processes are presented in Table~\ref{Table1}.  
 
\begin{figure}  
\centering  
\vspace{-0.2cm}  
\includegraphics[width=10.5cm,clip]{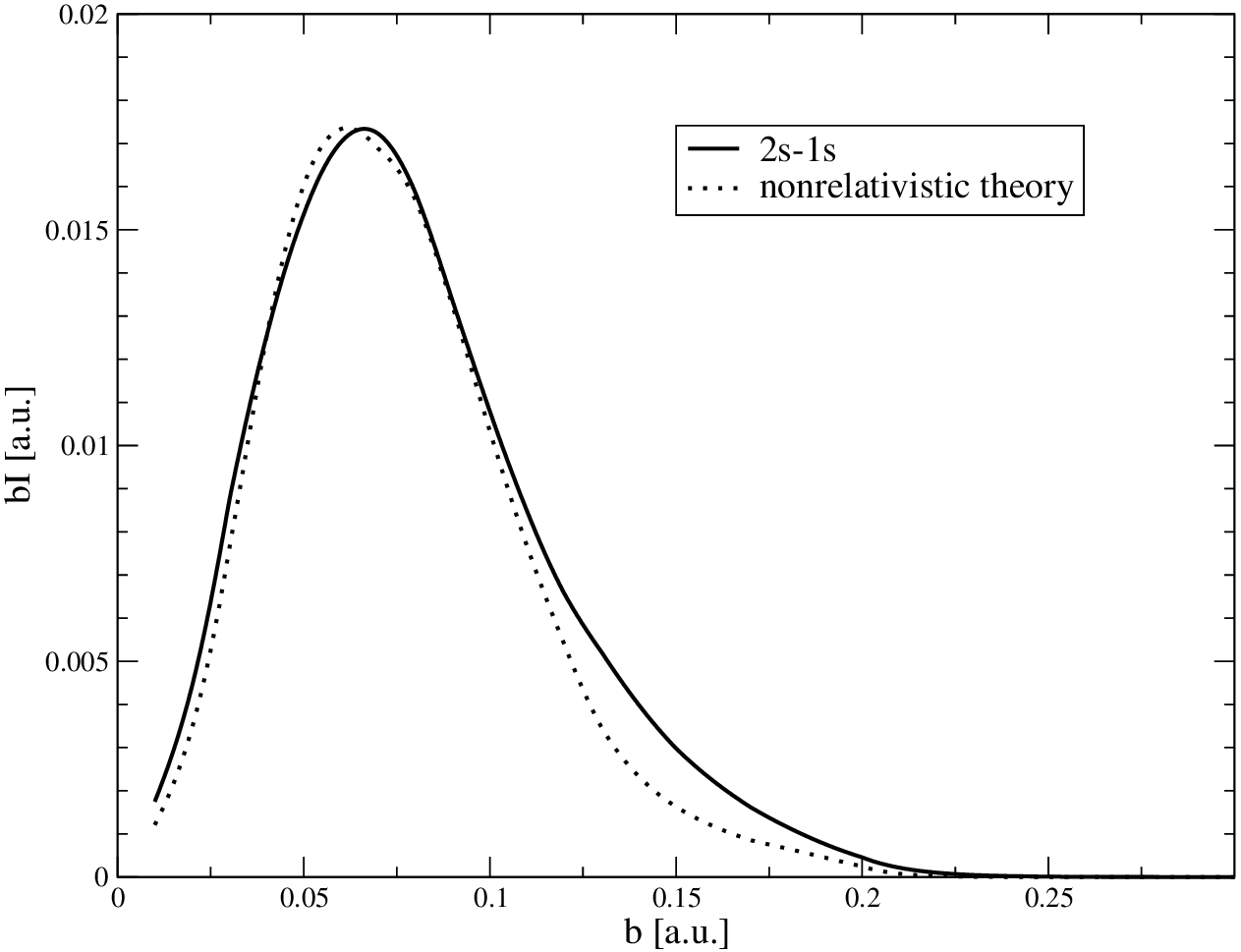}  
\caption{The intensity $I$ of the Bi K$^{''}_{\alpha_2}$ radiation   
weighted by the impact parameter in the Xe-Bi$^{83+}$ collision as a function   
of the impact parameter $b$. The dotted line indicates the results of   
the nonrelativistic calculations.}  
\label{Fig:Bi_L1rad_b}  
\end{figure}

\begin{figure}  
\centering  
\vspace{-0.2cm}  
\includegraphics[width=10.5cm,clip]{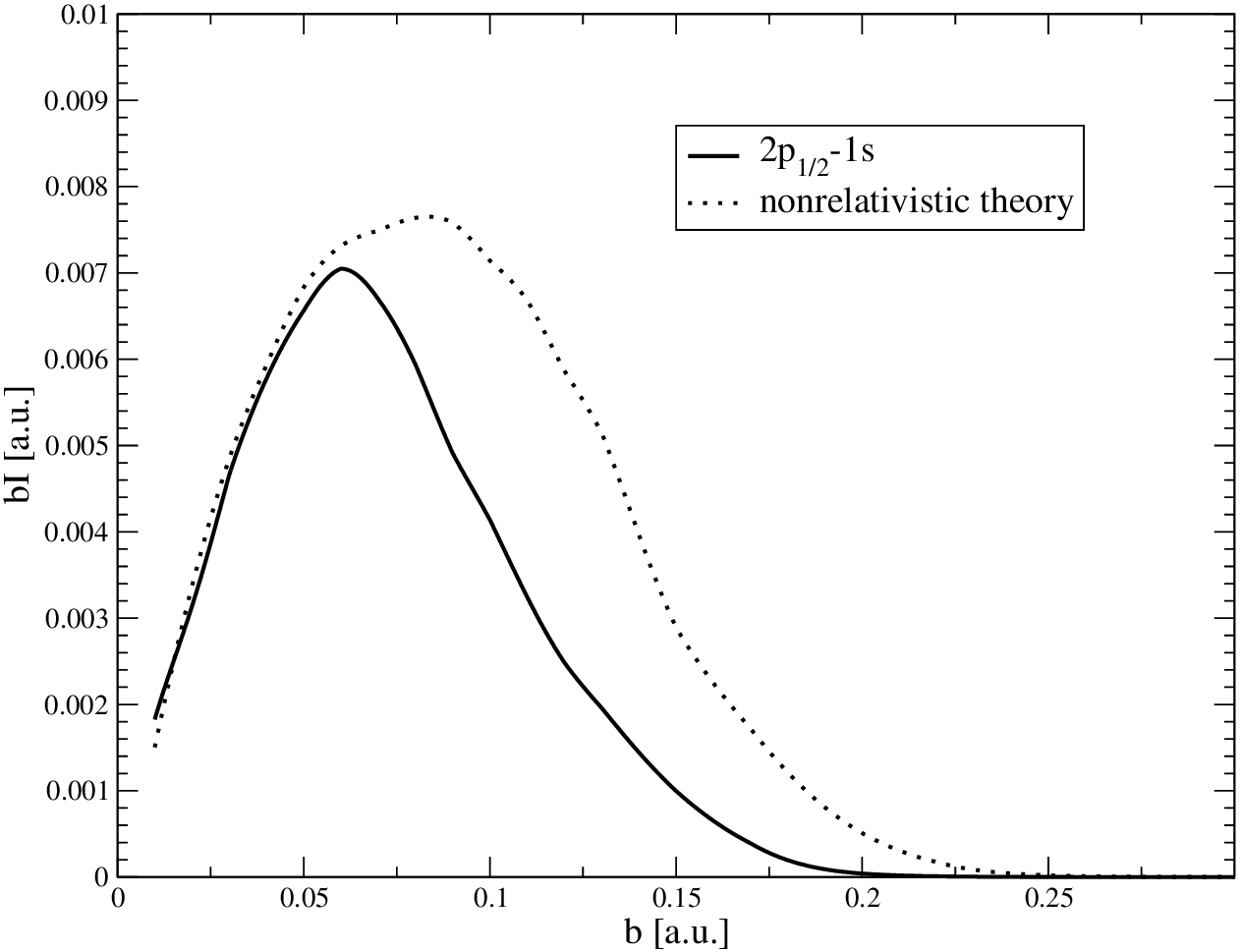}  
\caption{The intensity $I$ of the Bi K$^{'}_{\alpha_2}$ radiation   
weighted by the impact parameter in the Xe-Bi$^{83+}$ collision as a function   
of the impact parameter $b$. The dotted line indicates the results of   
the nonrelativistic calculations.}  
\label{Fig:Bi_L2rad_b}  
\end{figure}

\begin{figure}  
\centering  
\vspace{-0.2cm}  
\includegraphics[width=10.5cm,clip]{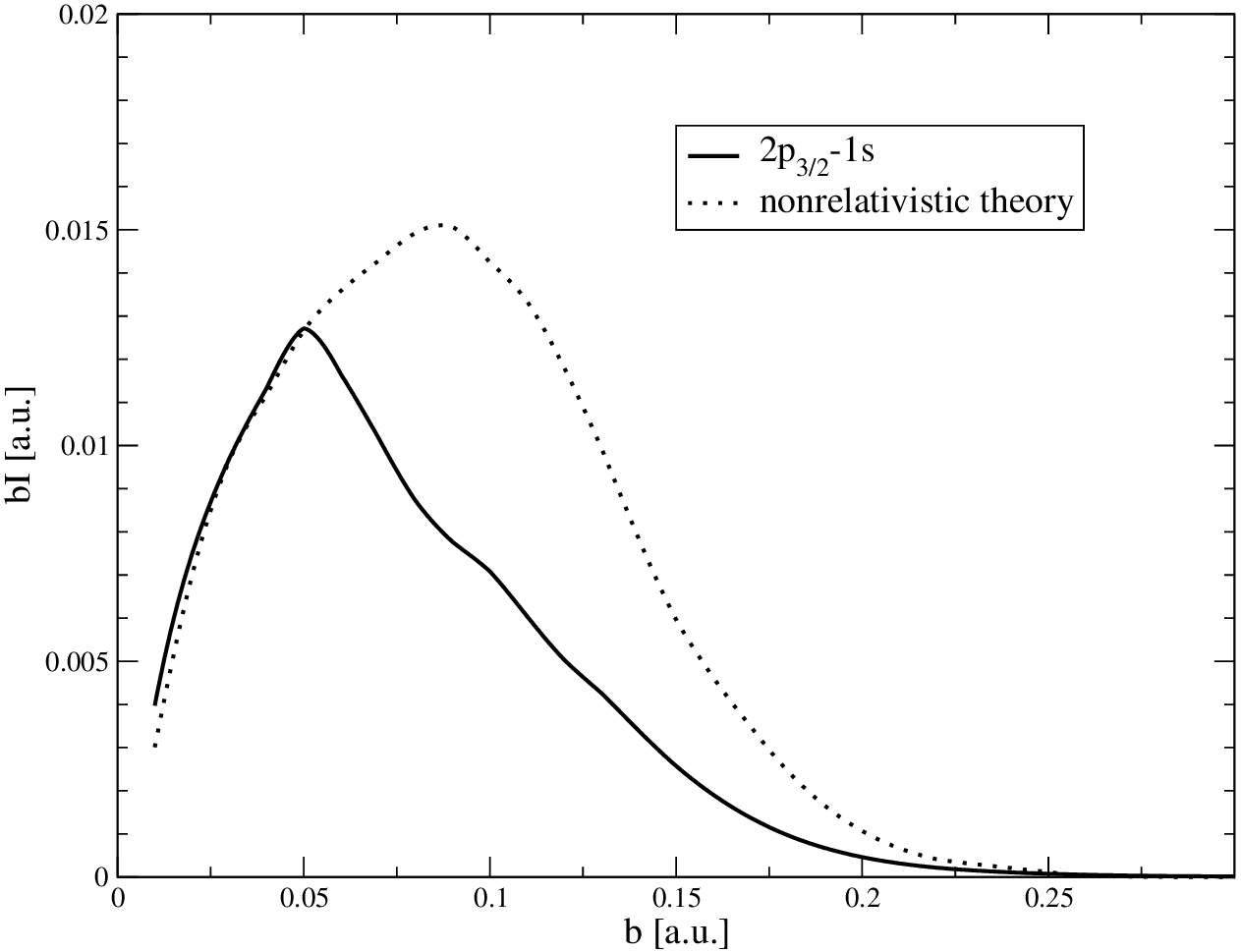}  
\caption{The intensity $I$ of the Bi K$_{\alpha_1}$ radiation   
weighted by the impact parameter in the Xe-Bi$^{83+}$ collision as a function   
of the impact parameter $b$. The dotted line indicates the results of   
the nonrelativistic calculations.}  
\label{Fig:Bi_L3rad_b}  
\end{figure}

According to our study (see section~\ref{subsec:Collision2}), the   
bismuth ion captures just a few electrons.  
Due to the small electron number and the rather large nuclear   
charge number, the probability of the Auger processes in the bismuth ion is 
negligible. In accordance with the values of the radiative transition   
probabilities~\cite{Jitrik:jpcrd:2004}, the electrons de-excite from the L$_1$,  
L$_2$, L$_3$, M, N, and other shells to the   
K shell  in the following order: first from the L$_2$- and L$_3$-shells,  
then from the M, N, and higher shells and, finally, from the L$_1$ shell.   
We also note that the probability to find two K-shell vacancies exceeds $0.8$  
for the impact parameter larger than $0.025$~a.u., where the K$_\alpha$   
radiation cross section is assembled (see below).  
But the maximum of the total probability to capture more than two electrons to  
the L, M, and N shells is less than $0.2$.   
That is why we assume that surely almost all the L$_2$-, L$_3$-, and even   
majority of the L$_1$-shell electrons de-excite radiatively to the K shell.   
Moreover, since the probability of the radiative de-excitation from the 
$np_{1/2}$ and  $np_{3/2}$ states to the $1s$ state by cascades is at least five 
times less than by the direct transition~\cite{Jitrik:jpcrd:2004} and the
population of the M, N and higher shells is low, the cascades 
processes leading to the K$_\alpha$ radiation can be neglected to a good 
accuracy.  
In order to obtain the intensities of the $2p_{3/2}$-$1s$ (K$_{\alpha_1}$),   
$2p_{1/2}$-$1s$ (K$_{\alpha_2}^{'}$), and $2s$-$1s$ (K$_{\alpha_2}^{''}$)  
lines, we calculate the probabilities $P_f$ of all possible atomic   
state configuration with $n_{K}$, $n_{L_1}$, $n_{L_2}$, $n_{L_3}$ electrons in  
the K, L$_1$, L$_2$, L$_3$ shells, correspondingly.  
Then we evaluate the radiative de-excitation probabilities    
with $m$ K$_{\alpha_1}$, K$_{\alpha_2}^{'}$ or K$_{\alpha_2}^{''}$ photons   
$P^{\rm{rad}}_{m, \rm{K}_{\alpha_i}}(f)$  
in accordance with the radiative transition probabilities from the L$_i$ to   
K shell for hydrogenlike bismuth $A_{\rm{L}_i}$   
($A_{\rm{L}_1}/A_{\rm{L}_2}=0$, $A_{\rm{L}_1}/A_{\rm{L}_3}=0$,   
$A_{\rm{L}_2}/A_{\rm{L}_3}=31/27$)~\cite{Jitrik:jpcrd:2004}.  
Finally, the relative x-ray radiation intensities are given by  
$I^{\rm{rad}}_{\rm{K}_{\alpha_i}}(b)=\sum_{f} m P^{\rm{rad}}_{m,   
\rm{K}_{\alpha_i}}(f) P_f(b) $.  
The results of the calculations of   
$I^{\rm{rad}}_{\rm{K}^{"}_{\alpha_2}}(b)$,   
$I^{\rm{rad}}_{\rm{K}^{'}_{\alpha_2}}(b)$, and   
$I^{\rm{rad}}_{\rm{K}_{\alpha_1}}(b)$   
weighted by the impact parameter are    
presented in Figures~\ref{Fig:Bi_L1rad_b},    
~\ref{Fig:Bi_L2rad_b}, and~\ref{Fig:Bi_L3rad_b}, respectively.   
In the figures, the obtained nonrelativistic data are indicated by the dotted   
lines. We note a rather strong influence of the relativistic effects on the   
intensities of the K$_{\alpha_1}$ and K$_{\alpha_2}^{'}$ (but not   
K$_{\alpha_2}^{"}$) lines.   
 
The final values of the radiative intensities per collision or the total cross  
sections of the radiation processes are collected in Table~\ref{Table1}.  
The uncertainties of the obtained data, being estimated rather conservatively,  
account for the errors due to both collision calculation and   
post-collisional analysis. For bismuth the uncertainty is   
much larger than for xenon. This is caused by the use of the moving  
orbitals and the more diffucult post-collisional analysis for the bismuth ion.  
The relativistic effects are really large and reach $50\%$ for   
the K$_{\alpha_1}$ and K$_{\alpha_2}^{'}$ intensities.  
 
The relative intensities and comparison with the   
experiment~\cite{Gumberidze:2011} are   
presented in Table~\ref{Table2}. It can be seen that  
the theoretical results are in a reasonable agreement with the experimental 
ones. Some underestimation of the theoretical intensities for the bismuth 
K$_{\alpha_1}$ emission might be explained by disregarding the cascade 
radiation processes from the higher-excited bismuth shells and a possible 
anisotropy of the K$_{\alpha_1}$ radiation (see, e.g., 
Ref~\cite{Surzhykov:prl:2002}; the X-ray spectrum in the experiment was 
recorded by a detector mounted at the observation angle of 150 deg). 
In the same way, the somewhat smaller experimental intensity of the Xenon 
L-radiation might possibly be due to the transitions from high $n$ states, 
which can not be fully resolved in the experiment.
We expect that the theoretical results can be further improved by the  
calculations with a larger basis set  and by including higher shells of bismuth 
to the post-collisional analysis.  
 
%
 
%
%
%
%
%

\begin{table}  
\caption{Cross sections   
$\sigma$ ($10^{-14}$ cm$^2$)   
of the x-ray radiation processes  
for the Xe-Bi$^{83+}$ collision.}  
\begin{center}  
 
\begin{tabular}{|c|c|c|c|c|c|}  
\hline   
Process&   
(Xe, K)&(Xe, L)&(Bi, K$_{\alpha_1}$)&  
(Bi, K$_{\alpha_2}^{'}$)&  
(Bi, K$_{\alpha_2}^{''}$)\\  
& & &($2p_{3/2}$-$1s$) &($2p_{1/2}$-$1s$)& ($2s$-$1s$)\\  
\hline  
 
$\sigma$ of the x-ray radiation& 47(3) & 200(25) & 20(6) & 13(4) & 26(10) \\  
Nonrelativistic theory           & 50    & 218     &  31      & 20    & 24       
\\  
 
\hline  
\end{tabular}  
\end{center}  
\label{Table1}  
\end{table}  
 
\begin{table}  
\caption{Relative intensities of the x-ray radiation for the Xe-Bi$^{83+}$   
collision.}  
\begin{center}  
 
\begin{tabular}{|c|c|c|c|}  
\hline   
&(Xe, L)$/$(Xe, K)  
&(Bi, K$_{\alpha_1}$)$/$(Xe, K)  
&(Bi, K$_{\alpha_2}$)$/$(Xe, K)  
\\  
\hline  
Theory [this work]& 4.2(6) & 0.43(14) & 0.83(30) \\  
Experiment \cite{Gumberidze:2011} & 3.6(2) & 0.59(3) & 0.69(3)  \\  
\hline  
\end{tabular}  
\end{center}  
\label{Table2}  
\end{table}  
 
\section{Conclusion}  
\label{subsec:Conclusion}  
We have evaluated the post-collisional x-ray   
emission in the Bi$^{83+}$ - Xe collision at the projectile energy 70 MeV/u,  
treating separately the collisional and post-collisional processes.   
The many-electron excitation, ionization and charge-transfer   
probabilities were calculated within the independent particle   
model using the coupled-channel approach with the atomic-like Dirac-Fock-Sturm  
orbitals.   
The inner-shell atom/ion processes were comprehensively studied and   
the corresponding probabilities were presented as functions of the impact   
parameter. The analysis of the post-collisional processes 
leading to the x-ray emission was based on the fluorescence yields, the 
radiation and Auger decay   
rates. It allows us to derive the x-ray radiation intensities  
and compare them with the related experimental data. The higher accuracy of the  
theoretical predictions for the xenon (target) x-ray radiation   
intensities is caused by  
the easier decay analysis and the use of the nonmoving  
atomic orbitals.  
The obtained results are in a reasonable agreement with the experimental data.  
 
The theoretical study demonstrates a very significant role of the relativistic  
effects, up to $50\%$ for the bismuth x-ray radiation intensities. Thus,   
investigations of heavy highly charged ion-atom collisions seem very   
promising for tests of relativistic and QED effects in  
scattering processes.

%
\clearpage  
\section{Acknowledgments}  
This work was supported by RFBR (Grants No.~14-02-31418, No.~13-02-00630,
No.~12-03-01140-a, and No.~14-02-00241), 
GSI, and SPbSU  (Grants No.~11.50.1607.2013, No.~11.38.269.2014, and 
No.~11.38.261.2014). Y.S.K. acknowledges the 
financial support from the Helmholtz Association and SAEC “Rosatom”.
The work of A.G. was supported by the Alliance Program of the Helmholtz 
Association (HA$216$/EMMI).
%
\clearpage  
  
%
\end {document}